\documentclass[aps,prl,preprint,nopacs,superscriptaddress]{revtex4}

\usepackage{epsf}
\usepackage{graphicx}
\usepackage{sidecap}
\usepackage{amsmath,amsfonts,amssymb}

\begin{document}

\title{Unconventional transformation of spin Dirac phase across a topological quantum phase transition}

\author{Su-Yang Xu}
\affiliation {Joseph Henry Laboratory, Department of Physics, Princeton University, Princeton, New Jersey 08544, USA}

\author{Madhab Neupane}
\affiliation {Joseph Henry Laboratory, Department of Physics, Princeton University, Princeton, New Jersey 08544, USA}
\author{Ilya Belopolski}\affiliation {Joseph Henry Laboratory, Department of Physics, Princeton University, Princeton, New Jersey 08544, USA}
\author{Chang Liu}
\affiliation {Joseph Henry Laboratory, Department of Physics, Princeton University, Princeton, New Jersey 08544, USA}

\author{Nasser Alidoust}\affiliation {Joseph Henry Laboratory, Department of Physics, Princeton University, Princeton, New Jersey 08544, USA}
\author{Guang Bian} \affiliation {Joseph Henry Laboratory, Department of Physics, Princeton University, Princeton, New Jersey 08544, USA}

\author{Shuang Jia}\affiliation {Department of Chemistry, Princeton University, Princeton, New Jersey 08544, USA}
\affiliation {International Center for Quantum Materials, Peking University, Beijing 100871, China}

\author{Gabriel Landolt}\affiliation {Swiss Light Source, Paul Scherrer Institute, CH-5232, Villigen, Switzerland}\affiliation {Physik-Institute, Universitat Zurich-Irchel, CH-8057 Zurich, Switzerland}
\author{Bartosz Slomski}\affiliation {Swiss Light Source, Paul Scherrer Institute, CH-5232, Villigen, Switzerland}\affiliation {Physik-Institute, Universitat Zurich-Irchel, CH-8057 Zurich, Switzerland}
\author{J. Hugo Dil}\affiliation {Swiss Light Source, Paul Scherrer Institute, CH-5232, Villigen, Switzerland}\affiliation {Physik-Institute, Universitat Zurich-Irchel, CH-8057 Zurich, Switzerland}

\author{Pavel P. Shibayev}
\affiliation {Joseph Henry Laboratory, Department of Physics, Princeton University, Princeton, New Jersey 08544, USA}

\author{Susmita Basak}
\affiliation {Department of Physics, Northeastern University, Boston, Massachusetts 02115, USA}

\author{Tay-Rong Chang}
\affiliation{Department of Physics, National Tsing Hua University, Hsinchu 30013, Taiwan}
\author{Horng-Tay Jeng}\affiliation {Department of Physics, National Tsing Hua University, Hsinchu 30013, Taiwan}\affiliation {Institute of Physics, Academia Sinica, Taipei 11529, Taiwan}

\author{Robert J. Cava}\affiliation {Department of Chemistry, Princeton University, Princeton, New Jersey 08544, USA}
\author{Hsin Lin}
\affiliation {Graphene Research Centre and Department of Physics, National University of Singapore 11754, Singapore}
\author{Arun Bansil} \affiliation {Department of Physics, Northeastern University, Boston, Massachusetts 02115, USA}
\author{M. Zahid Hasan} \affiliation {Joseph Henry Laboratory, Department of Physics, Princeton University, Princeton, New Jersey 08544, USA}\affiliation{Princeton Center for Complex Materials, Princeton Institute for Science and Technology of Materials, Princeton University, Princeton, New Jersey 08544, USA}

\begin{abstract}

The topology of a topological material can be encoded in its surface states. These surface states can only be removed by a bulk topological quantum phase transition into a trivial phase. Here we use photoemission spectroscopy to image the formation of protected surface states in a topological insulator as we chemically tune the system through a topological transition. Surprisingly, we discover an exotic "pre-formed" spin-momentum locked, gapped surface state in the trivial phase that shares many important properties with the actual topological surface state in anticipation of the change of topology. Using a spin-resolved measurement, we show that apart from a surface band-gap these "pre-formed" states develop spin textures similar to the topological surface states well-before the transition. Our results offer a general paradigm for understanding how surface states in topological phases arise and are suggestive for future realizing Weyl arcs, condensed matter supersymmetry and other fascinating phenomena in the vicinity of topological quantum criticality.

\end{abstract}
\date{\today}
\maketitle

Understanding the physics of distinct phases of matter is one of the most important goals in physics in general. For a new phase of matter, a powerful route toward such understanding is to study the way it arises from an understood state by investigating the nature of a phase transition. A topological insulator (TI), a weakly interacting electronic system, is a distinct phase of matter that cannot be adiabatically connected to a conventional material without going through a topological quantum ($T\rightarrow0$ K) phase transition (TQPT), which involves a change of the bulk topological invariant without invoking any many body interaction. The discovery of the 3D Z$_2$ topological insulator state has attracted huge interest and led to a surge of research in finding new and engineered topological states \cite{SUSY, Senthil, Moore, Zahid RMP, TI_book_2014, Zhang RMP, Kane_PRL, Pallab, David Nature BiSb, David Nature tunable, TCI_Story, Hasan TCI, TCI_Ando, Coleman, MN, Shi, Feng, 3D_Dirac, CdAs_Hasan, CdAs_Cava, Chen_Na3Bi, Hasan_Na3Bi, Ashvin, Gil, Qi}. Many new topological phases of matter, such as a topological crystalline insulator \cite{TCI_Story, Hasan TCI, TCI_Ando}, a topological Kondo insulator \cite{Coleman, MN, Shi, Feng}, a topological Dirac/Weyl semimetal \cite{3D_Dirac, Pallab, CdAs_Hasan, CdAs_Cava, Chen_Na3Bi, Hasan_Na3Bi, Ashvin}, etc. have just been predicted or realized. All these phases are predicted to feature protected surface states, which serve as the experimental signature for their nontrivial topology in the bulk, and they are in fact formed via TQPTs and need to be understood in real materials. Therefore, it is of general importance to study how protected surface state emerge from a trivial material by crossing the topological-critical-point (TCP) of a TQPT. However, to date, the electronic and spin groundstate in the vicinity of a TCP for any topological systems remains elusive.

As an example, for a Z$_2$ topological insulator, it is well established that the odd number of Dirac surface states and their spin-momentum locking are the signature that distinguishes it from a conventional insulator. However, an interesting and vital question that remains unanswered is how topological surface states emerge as a non-topological system approaches and crosses the TCP. The most straightforward answer is that there are neither surface states nor spin polarization in the conventional insulator (non-topological) regime. In this case, the gapless topological surface states and spin-momentum locking set in abruptly and concomitantly at the topological-critical-point. However, there might be more exotic scenarios. Therefore an experimental study focused on this topic is needed to settle this issue.

Understanding the nature of a TCP is also of broad interest because recent theories have proposed a wide range of exciting quantum phenomena based on topological criticality. It has been proposed that the TCP of various TQPTs can not only realize new groundstates such as higher dimensional Dirac fermions \cite{3D_Dirac, Pallab}, Weyl fermions under magnetization \cite{3D_Dirac, Pallab, Gil, Qi}, supersymmetry SUSY state \cite{SUSY} and interacting topological states \cite{Senthil}, but also show exotic transport and optical responses such as chiral anomaly in magnetoresistance \cite{Chiral} or the light-induced Floquet topological insulator state \cite{Galitski}. In order to achieve them in real materials, it is also quite suggestive to study the electronic and spin groundstate in the vicinity of the TCP in some great detail.


In this article, we report the observation of an exotic phenomenon associated with the formation of the topological surface states across a TQPT. We show that there exists spin-momentum locked but gapped surface states on the topologically trivial side of the TQPT that serves as novel precursor states to the topological surface states. These surface states are systematically enhanced and evolve into the actual topological surface states across the TCP. This is particularly interesting because it can be viewed as a novel proximity effect due to the adjacent topological insulator phase. To achieve these, we systematically study the evolution of electronic and spin groundstate near the topological-critical-point with a step finer than 2\% in the prototypical TQPT BiTl(S$_{1-\delta}$Se$_\delta$)$_2$ system. The BiTl(S$_{1-\delta}$Se$_\delta$)$_2$ system is known to host one of the most basic TQPTs between a conventional band insulator and a 3D Z$_2$ topological insulator (TI) \cite{Suyang, Sato, Oh}, and is therefore an ideal platform for our goal. We show that even though the bulk material of BiTl(S$_{1-\delta}$Se$_\delta$)$_2$ lies in the conventional semiconductor regime, we observe an unexpected gapped quasi-two-dimensional electron gas that shares many properties with actual topological surface states. Surprisingly, our critical spin-resolved data reveal that these gapped states carry spin polarization, whose momentum space texture at the native Fermi level resembles that of on the surface of a topological insulator. We further show that the observed spin-textured surface states prominently dominate the surface low-energy physics upon approaching the TCP, and systematically evolve into the gapless topological surface states. Our observation sets a general paradigm for understanding how topological surface states can arise from a conventional material by going through a TQPT, which is of value for studying various new topological phases and the formation of their protected surface states \cite{TCI_Story, Hasan TCI, TCI_Ando, Coleman, Pallab, MN, Shi, Feng, 3D_Dirac, CdAs_Hasan, CdAs_Cava, Chen_Na3Bi, Hasan_Na3Bi, Ashvin, Gil, Qi}. The gapped spin-helical surface states also suggest the remarkable potential for the utilization of unique gapped spin-textured electrons on the surface of a carefully designed conventional semiconductor using spin-polarized tunneling or band-selective optical methods in future applications.

\bigskip
\bigskip
\textbf{Results}
\newline
\textbf{Spin-integrated electronic structure near the TCP}
\newline
We present in-plane electronic structure ($E_{\textrm{B}}$ vs $k_{\|}$) of the BiTl(S$_{1-\delta}$Se$_\delta$)$_2$ system at varying compositions ($\delta$). Fig. 1 shows that the two end compounds ($\delta=0.0$ and $1.0$) are in clear contrast, namely, $\delta=0.0$ has no surface states and $\delta=1.0$ has surface states connecting the bulk conduction and valence bands, which clearly reveals the difference between the conventional semiconductor phase and the Z$_2$ topological band insulator phase, in agreement with previous studies \cite{Suyang, Sato}. The conventional semiconductor state is found to extend from $\delta=0.0$ to $0.4$ (Fig. 1a), whereas the topological state is clearly observed from $\delta=1.0$ to $0.6$ (Fig. 1c). A small but observable bulk band-gap of about 30 meV is observed for $\delta=0.45$ in Fig. 1e, indicating that the system continues to belong to the conventional semiconductor phase. Upon increasing $\delta$ to the region of $0.475 - 0.525$, the bands are found to further approach each other, and the linear dispersion behavior of the bands is observed to persist at energies all the way across the node (the Dirac point). Thus based on the observed linear dispersion, the critical composition can be estimated to be $\delta_c=0.5\pm0.03$. At $\delta=0.60$ (Fig. 1c), a clear bulk conduction band is observed inside the surface states' upper Dirac cone. Moreover, the bulk conduction and valence bands are separated by an observable bulk gap, which is traversed by the gapless topological surface states. Thus, our data show that the system belongs to the topological insulator regime for compositions of $\delta\geq0.60$. As for the system lying very close to the bulk inversion at $\delta=0.50$ or $0.525$, based on the in-plane dispersion data in Fig. 1b alone, the nature of the observed Dirac-like band cannot be conclusively determined, because it can be interpreted as two-dimensional topological surface states or three-dimensional bulk Dirac states \cite{Pallab} expected near the bulk band inversion. However, one of the two possibilities can be identified by measuring the dispersion along the out-of-plane $k_z$ direction, since the three-dimensional bulk Dirac states are expected to be highly dispersive \cite{Suyang, CdAs_Hasan, Chen_Na3Bi, Hasan_Na3Bi} (the velocity along $k_z$ direction of the 3D bulk Dirac band at $\delta_c$ is estimated to be $\sim2.5$ eV$\cdot{\textrm{\AA}}^{-1}$ \cite{Suyang}), whereas the 2D surface states are not expected to show observable dispersion along the $k_z$ direction.

Thus, in order to better understand the nature of the bands at compositions near the TCP, we perform angle-resolved photoemission spectroscopy (ARPES) measurements as a function of incident photon energy values (Fig. 1e,f) to probe their out-of-plane $k_z$ dispersion. Upon varying the photon energy, one can effectively probe the electronic structure at different out-of-plane momentum $k_z$ values in a three-dimensional Brillouin zone. Left panel of Fig. 1f shows the incident photon energy ($k_z$) measurements at $\delta=0.525$ by the Fermi surface mapping in $k_\|$ vs $k_z$ momentum space. The straight Fermi lines that run parallel to the $k_z$ axis show nearly absence of observable $k_z$ dispersion. Similarly, incident photon energy measurements are also performed at compositions $\delta=0.40$ and $0.45$, where a clear bulk band-gap is observed ($\nu_0=0$). Surprisingly, even for the gapped electronic structure at $\delta=0.40, 0.45$, our data show clear absence of $k_z$ dispersion. Therefore the observed bands cannot be interpreted as three-dimensional bulk Dirac bands expected near the bulk band inversion. In fact, our systematic $k_z$ measurements (see Supplementary Figure 1 for more data from 14 eV to 70 eV) reveal that the electronic states near the Fermi level of the $\delta=0.40, 0.45$ samples are two-dimensional. Due to the coexistence of bulk states at the same energy, it is most precise to name them as quasi-2D states. But because they are strongly localized near the surface of the sample and because they smoothly evolve into the topological surface states as the system is tuned across the TQPT, we believe that it is also reasonable to call them as surface states. Such anomalously strong surface states on the trivial side suggest that these states are due to their proximity to the topological insulator regime.


\bigskip
\bigskip
\textbf{Spin-resolved measurements below and close to the TCP}
\newline
In order to study the spin properties of the observed anomalous surface states, we perform spin-resolved measurements on the system with compositions below and near the TCP. We present spin-resolved data taken on the composition of $\delta=0.40$ (Fig. 2a) and focus on the vicinity of the Fermi level ($E_{\textrm{B}}=-0.02$ eV). The momentum distribution curves (MDCs) for the spectrum are shown in Fig. 2b, where the highlighted curve is chosen for spin-resolved (SR) measurements. Figs. 2d-g shows the in-plane SR-MDC spectra as well as the measured in-plane spin polarization along the $\bar{\Gamma}$-$\bar{M}$ and $\bar{\Gamma}$-$\bar{K}$ momentum space cuts. Clear in-plane spin polarization is observed on the surface states from Figs. 2d-g. Furthermore, the measured spin polarization in Figs. 2d-g shows that the spin texture is arranged in a way that spins have opposite directions on the opposite sides of the Fermi surface. In addition, the out-of-plane component of the spin polarization along the $\bar{\Gamma}$-$\bar{M}$ and $\bar{\Gamma}$-$\bar{K}$ cuts is shown in Figs. 2h-k. No significant out-of-plane spin polarization (Figs. 2h-k) is observed within our experimental resolution. The spin texture configuration can be obtained from the spin-resolved measurements in Figs. 2d-k, as schematically shown by the arrows in Fig. 2c. Surprisingly, our spin-resolved measurements reveal that these surface states are not only strongly spin-polarized, but their spin texture near the native Fermi level resembles the helical spin texture on the topological surface states as observed in Bi$_2$Se$_3$ \cite{David Nature tunable}.

We present systematic spin-resolved studies to understand the way spin texture of the surface states evolves as a function of binding energy $E_{\textrm{B}}$ and composition $\delta$. Figs. 3a-d show spin-resolved data at different binding energies for a sample with $\delta=0.40$. The spin-momentum locking behavior is observed at all binding energies from near the Fermi level ($E_{\textrm{B}}=-0.02$ eV) to an energy near the conduction band minimum ($E_{\textrm{B}}=-0.32$ eV). While the magnitude of the spin polarization on the Fermi level is found to be around 0.3, the spin polarization magnitude is found to decrease to nearly zero while approaching small values of momenta near the Kramers' point $\bar{\Gamma}$ (the conduction band minimum). Furthermore, at energies cutting across the bulk valence band at $E_{\textrm{B}}=-0.57$ eV, $E_{\textrm{B}}=-0.72$ eV, the measured spin polarization profile is clearly reversed, where a right-handed profile is found for the surface states on the boundary. In addition, the magnitude of the spin polarization is found to be increased as the energy is tuned away from the bulk band-gap, which is consistent with the gapped nature of the surface states. The observed reduction of net spin polarization at small momenta and the absence of net spin polarization at the $\bar{\Gamma}$ ($k=0$, see Supplementary Figure 2) point are important for the gapped nature of surface states in $\delta=0.4$ samples. As for the gapless case with the system composition at $\delta=0.50$, the spin-resolved measurements (Figs. 4a-d) reveal the same helical-like spin texture configuration on the Fermi level, where the magnitude of the spin polarization is around 0.5 at the Fermi level in this composition. However, in contrast to the $\delta=0.4$ case, it does not show any obvious reduction in going to small values of momenta near the Kramers' point $\bar{\Gamma}$ (spin polarization $\sim$ 0.45 for $E_{\textrm{B}}=-0.32$ eV), which is consistent with its gapless nature. The adequate energy-momentum resolution of our SR-ARPES instrument in order to resolve opposite spins at small momenta, such as $k\sim0.05$ $\textrm{\AA}^{-1}$, is demonstrated by these SR measurements on $\delta=0.50$, which strongly supports that the observed strong spin polarization reduction at the $\delta=0.40$ case reveals an intrinsic property of the system relevant to the topological transition. Finally, we present the spin data taken on the composition far into the topologically trivial side ($\delta=0.0$). Our spin-resolved measurements (Figs. 4e-h) show only very weak polarizations ($\sim 0.05$), which lie within the uncertainty levels of the measurements. The magnitude of the spin polarization is too weak (comparable to the instrumental resolution) to obtain the spin texture configuration around the Fermi surface for samples with $\delta=0.0$. The observed weak polarization on $\delta=0.0$ suggests that the surface states are much suppressed in going away from the TCP (such as the $\delta=0.0$ samples). More systematic spin-resolved studies can be found in Supplementary Figures 2,3.

In Supplementary Discussion and Methods, we model the surface of a topological phase transition system based on the $4\times4$ $k \cdot p$ model \cite{kdotp} and utilize the Green's function method to obtain the spectral weight as well as the spin polarization near the surface region of the system as a function of bulk band-gap value in the model. We found a reasonable qualitative agreement between our experimental results and the $k \cdot p$ model calculation as seen in Supplementary Figure 4.

\bigskip
\bigskip
\textbf{Discussion}
\newline

Although the observed surface states share important properties with actual topological surface states, the following observations from our data clearly show that they are still consistent with the non-topological bulk regime. First, the experimentally observed surface states are gapped, and disperse roughly along the edge of the bulk continuum. Thus they do not connect or thread states across the bulk band-gap as in a Z$_2$ topological insulator. Second, it is also possible to choose an energy value within the bulk band-gap for samples lying in the conventional semiconductor regime (e.g. $\delta=0.4$), so that no surface state is traversed, consistent with the topological triviality of the sample. These experimental facts guarantee that the observed surface states at $\delta\lesssim0.5$ are consistent with the conventional semiconductor phase of the system ($\nu_0=0$, trivial Z$_2$ index). In the Supplementary Discussion and Supplementary Figures 5,6, we propose a phenomenological picture that involves a Rashba-like state for our observed spin-momentum locked surface states on the trivial side prior to the TQPT. Nevertheless, the detailed theoretical understanding requires a microscopic theory to show why spin-momentum locking is formed even before the TQPT actually takes place. Regardless of the theoretical studies required in the future, our observations experimentally reveal a novel proximity effect due to the adjacent TI phase in a TQPT system. First, the surface states on the trivial side show spin polarization texture that resembles a topological insulator. Second, their surface spectral weight and magnitude of spin polarization are enhanced as we approach the TCP. Third, they evolve into the topological surface states. These measurements with spin and momentum resolution clearly show that these surface states are critically relevant to the bulk band inversion and TQPT in the bulk. In the Supplementary Figure 7, we show similar observation near the TCP of another prototypical TQPT system (Bi$_{1-\delta}$In$_{\delta}$)$_2$Se$_3$ \cite{Oh}. Therefore, these systematic and careful measurements on multiple systems suggest that our observation is unlikely a special case due to material details of the BiTl(S$_{1-\delta}$Se$_{\delta}$)$_2$ system but an important proximity phenomenon that describes the TCP in the electronic and spin groundstates in many TQPT systems. Our observation can also be applied to explain a number of recent experiments on some newly predicted topological matter, such as the topological Kondo insulator phase predicted in SmB$_6$ \cite{Coleman}. In SmB$_6$, the Kondo hybridization gap is believed to become significant below 30 K and the low temperature resistivity anomaly occurs below 6 K \cite{MN}. However, ARPES experiments have observed quasi-2D low energy states without $k_z$ dispersion persisting up to temperatures $\geq100$ K \cite{Feng, Shi}. Furthermore, a recent theoretical effort proposed that the formation of spin-momentum locked surface states prior to the TQPT is due to the reversal of bulk Dirac fermion chirality across the TQPT \cite{Vidya}, which is consistent with our systematic experimental data. We also note that the quantum fluctuation can be another interesting direction, which is widely studied in many non-topological quantum phase transitions, where a local order parameter is present. However, in order to study that in our system, it is necessary to first theoretically understand the role of quantum fluctuations in a topological quantum phase transition in a non-interacting or weakly-interacting system. Only then, it is possible to identify the correct experimental probe that can be sensitive to the quantum fluctuation in our studied system.

Irrespective of the theoretical origin, our observation itself is important for further considerations of these novel states. We propose potential device applications for the spin-textured gapped surface states that we observed. Since there is a true energy gap without any states (neither surface nor bulk) in the conventional insulator phase (such as the $\delta=0.4$ sample), thus in this case the spin-textured surface electrons can be turned on and off via tuning the chemical potential of the samples, which realizes a novel switch of the spin-textured surface electrons not possible for a usual topological surface state without adding magnetism. Such a spin switch is experimentally demonstrated in BiTl(S$_{1-\delta}$Se$_{\delta}$)$_2$ through NO$_2$ surface adsorption on a $\delta=0.4$ sample under UHV conditions as shown in Fig. 5.  Due to the fact that there are bulk bands at the same energy where these surface states exist, they need to be investigated and utilized using surface-sensitive approaches. For example, in STM measurements, they open up new possibilities for observing anomalous tunneling behavior, anomalous transmission near step edges and other unusual surface effects on a conventional semiconductor surface, which can be switched on and off via changing the sample bias in tunneling experiments. Similarly, our observations also enable band-selective optical experiments, such as photocurrent and photoconductivity manipulation using circularly polarized incident light \cite{Optical}, leading to potential opto-spintronics applications. Further research toward these goals will require higher quality nano-structured molecular beam epitaxy grown samples.

\bigskip
\bigskip
\textbf{Methods}
\newline
\textbf{Electronic structure measurements.}
Spin-integrated angle-resolved photoemission spectroscopy (ARPES) measurements were performed with incident photon energy of $8$ eV to $30$ eV at beamline 5-4 at the Stanford Synchrotron Radiation Lightsource (SSRL) in the Stanford Linear Accelerator Center, with $26$ eV to $90$ eV at beamlines 4.0.3, 10.0.1, and 12.0.1 at the Advance Light Source (ALS) in the Lawrence Berkeley National Laboratory (LBNL), and with $16$ eV to $50$ eV at the PGM beamline in the Synchrotron Radiation Center (SRC) in Wisconsin. Samples were cleaved in situ between $10$ to $20$ K at chamber pressure better than $5 {\times} 10^{-11}$ torr at all endstations at the SSRL, the ALS and the SRC, resulting in shiny surfaces. Energy resolution was better than 15 meV and momentum resolution was better than 1\% of the surface Brillouin zone (BZ). Adsorption of NO$_2$ molecules on sample surface was achieved via controlled exposures to NO$_2$ gas (Matheson, 99.5\%). The adsorption effects were studied under static flow mode by exposing the clean sample surface to the gas for a certain time at the pressure of $1 \times 10^{-8}$ torr, then taking data after the chamber was pumped down to the base pressure. Spectra of the NO$_2$ adsorbed surfaces were taken within minutes of opening the photon shutter to minimize potential photon induced charge transfer and desorption effects.

\textbf{Spin-resolved measurements.}
Spin-resolved ARPES measurements were performed on the SIS beamline at the Swiss Light Source (SLS) using the COPHEE spectrometer with two 40 kV classical Mott detectors and photon energy of 20 to 70 eV, which systematically measures all three components of the spin of the electron (P$_x$, P$_y$, and P$_z$) as a function of its energy and momentum \cite{Hugo Review}. Energy resolution was better than 60 meV and momentum resolution was better than 3\% of the surface BZ. Samples were cleaved in situ at 20 K at chamber pressure less than $2{\times}10^{-10}$ torr. Typical electron counts on the Mott detector reached $5\times10^5$, which placed an error bar of $\pm0.01$ for the data points in all spin polarization measurements. Our spin-resolved ARPES measurements were performed with linearly $p$-polarized light at synchrotron radiation energies 20 eV to 70 eV, where the final state effects are demonstrated to be negligible \cite{Rader}.

\textbf{Sample growth.}
Single crystals of BiTl(S$_{1-\delta}$Se$_{\delta}$)$_2$ were grown from high purity elements mixed in a stoichiometric ratio using the Bridgman method systematically described in Ref. \cite{Suyang, Jia}. The mixture was heated in a clean evacuated quartz tube to 900$^{\circ}$C where it was held for two days. Afterwards, it was cooled slowly, at a rate of $1.5^{\circ}\textrm{C}$ per hour in the vicinity of the melting point, from the high temperature zone towards room temperature. The spatial compositional homogeneity of the cleaved sample surfaces were confirmed using high resolution energy dispersive spectroscopy (EDS). The EDS measurements were performed at the Imaging and Analysis Center (IAC) at the Princeton University's Institute for the Science and Technology of Materials (PRISM). The equipment used was a FEI company Quanta 200 f field emission Environmental Scanning Electron Microscope (ESEM) equipped with an Oxford INCA EDS data analysis software system and an Oxford XMAX 80 mm$^2$ high efficiency EDS detector (see Supplementary Figure 8,9 and Supplementary Methods for details). We note that the $n$-doping in the samples were caused by Se or S vacancies \cite{David Nature tunable}. And depending on the number of vacancies in a crystal, its chemical potential slightly varied from batch to batch. We did notice a small variation of the chemical potential (e.g. see Fig. 1b of the main text). However, since the samples were always $n$-type, we could always observe the conduction band, the valence band, the band-gap, and the surface states in ARPES. Thus the small variation of the chemical potential did not affect our results and conclusions.

\textbf{First-principles and model theoretical calculation methods.}
The theoretical band calculations were performed with the LAPW method using the WIEN2K package \cite{Blaha} within the framework of density functional theory (DFT). The generalized gradient approximation was used to model exchange-correlation effects. For the semi-infinite surface system, model calculations were done based on a Green's function with implementation of the experimentally-based $k \cdot p$ model \cite{kdotp} to reveal the electronic structure and spin configuration near the surface region of the BiTl(S$_{1-\delta}$Se$_\delta$)$_2$ system.

\bigskip
\bigskip
\textbf{Acknowledgements}
\newline
Work at Princeton University is supported by U.S. DOE/BES DE-FG-02-05ER46200. M.Z.H. acknowledges visiting-scientist support from Lawrence Berkeley National Laboratory and additional support from the A. P. Sloan Foundation. The spin-resolved photoemission measurements using synchrotron X-ray facilities were supported by the Swiss Light Source, the Swiss National Science Foundation. Crystal growth was supported by NSF-DRM-0819860. S. Jia was supported by National Basic Research Program of China (Grant Nos. 2013CB921901 and 2014CB239302).Theoretical computations were supported by the US Department of Energy (DE-FG02-07ER46352 and AC03-76SF00098) as well as the National Science Council and Academia Sinica in Taiwan, and benefited from the allocation of supercomputer time at NERSC and Northeastern University's Advanced Scientific Computation Center. H.L. acknowledges the Singapore National Research Foundation for the support under NRF Award No. NRF-NRFF2013-03. T.R.C. and H.T.J. were supported by the National Science Council, Taiwan. We thank Sung-Kwan Mo, Alexei Fedorov, Jonathan Denlinger, and Makoto Hashimoto for beamline assistance at the LBNL and the SSRL. We also gratefully acknowledge Nan Yao and Gerald R. Poirier for the assistance of the EDS measurements at the Imaging and Analysis Center (IAC) at the Princeton University. The work also benefited from helpful discussions with Chen Fang.

\bigskip
\bigskip
\textbf{Author contributions}
\newline
S.-Y.X. and M.N. performed the experiments with assistance from I.B., C.L., N.A., G.B., P.P.V., and M.Z.H.; S.J. and R.J.C. provided samples; G.L., B.S., and J.H.D. provided beamline assistance; S.B., T.R.C., H.T.J., H.L., and A.B. carried out the theoretical calculations; M.Z.H. was responsible for the overall direction, planning and integration among different research units.

\bigskip
\bigskip
\textbf{Additional information}

The authors declare no competing financial interests. Correspondence should be addressed to M.Z.H. (Email: mzhasan@princeton.edu).

\newpage
\begin{figure*}
\centering
\includegraphics[width=17cm]{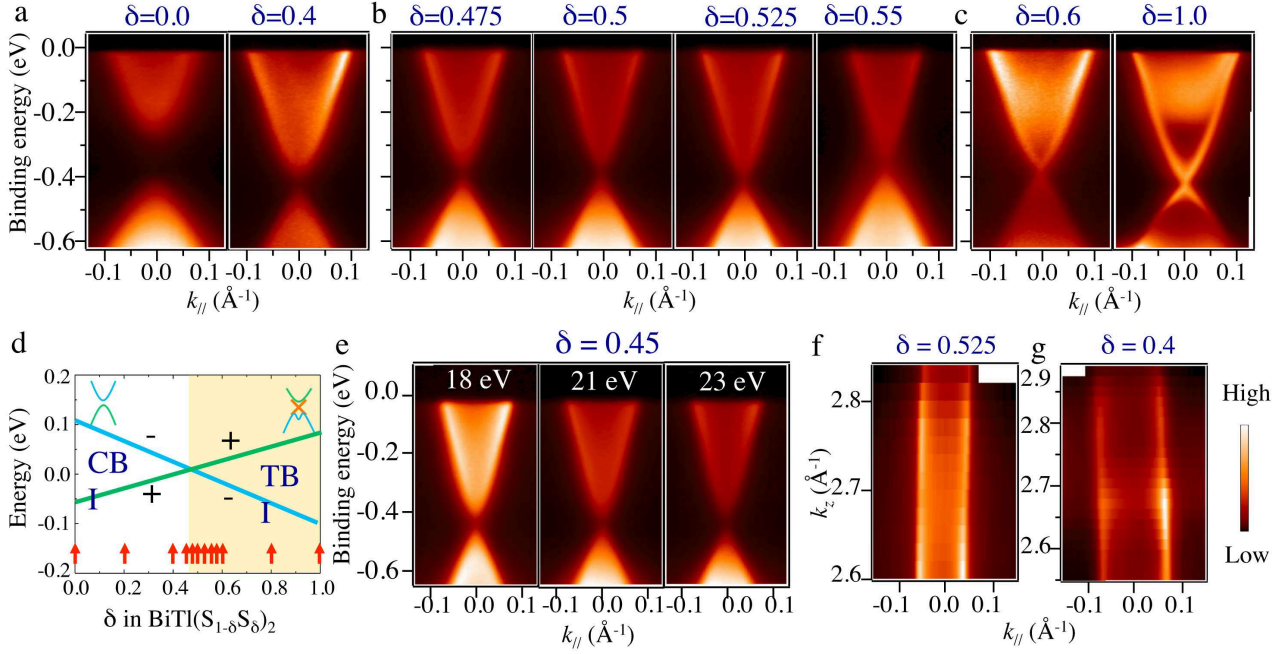}
\caption{\textbf{Observation of gapped quasi-two-dimensional states prior to the topological-critical-point of the topological quantum phase transition.} \textbf{a-c,} ARPES $k_\|$-$E_{\textrm{B}}$ maps of BiTl(S$_{1-\delta}$Se$_\delta$)$_2$ obtained using incident photon energy of 16 eV. The nominal composition values (defined by the mixture weight ratio between the elements before the growth) are noted on the samples. For \textbf{a,} conventional band insulator (CBI), a band-gap is clearly observed for $\delta$ = 0.0 to 0.4; For \textbf{b,} Compositions near the topological-critical-point (TCP) of the topological quantum phase transition (TQPT), $\delta$ = 0.45, 0.50, 0.525 and 0.55; And for \textbf{c,} topological band insulator (TBI), the conduction and valence bands are observed to be well-separated with the surface states connecting the band-gap for $\delta$ = 0.6 to 1.0. \textbf{d,} The energy levels of the first-principles calculated bulk conduction and valence bands of the two end compounds ($\delta=0.0$ and $1.0$) are connected by straight lines to denote the evolution of the bulk bands. The compositions selected for detailed experimental studies are marked by red arrows. The $+$ and $-$ signs represent the odd and even parity eigenvalues of the lowest lying conduction and valence bands of BiTl(S$_{1-\delta}$Se$_\delta$)$_2$. \textbf{e,} Incident photon energy dependence spectra for $\delta$ = 0.45.  \textbf{f,} $k_z$ vs $k_{\|}$ Fermi surface maps for $\delta$ = 0.525 and 0.4. The $k_z$ range shown for $\delta$ = 0.4 samples corresponds to the incident photon energy from 14 eV to 26 eV.}
\end{figure*}

\begin{figure*}
\centering
\includegraphics[width=17cm]{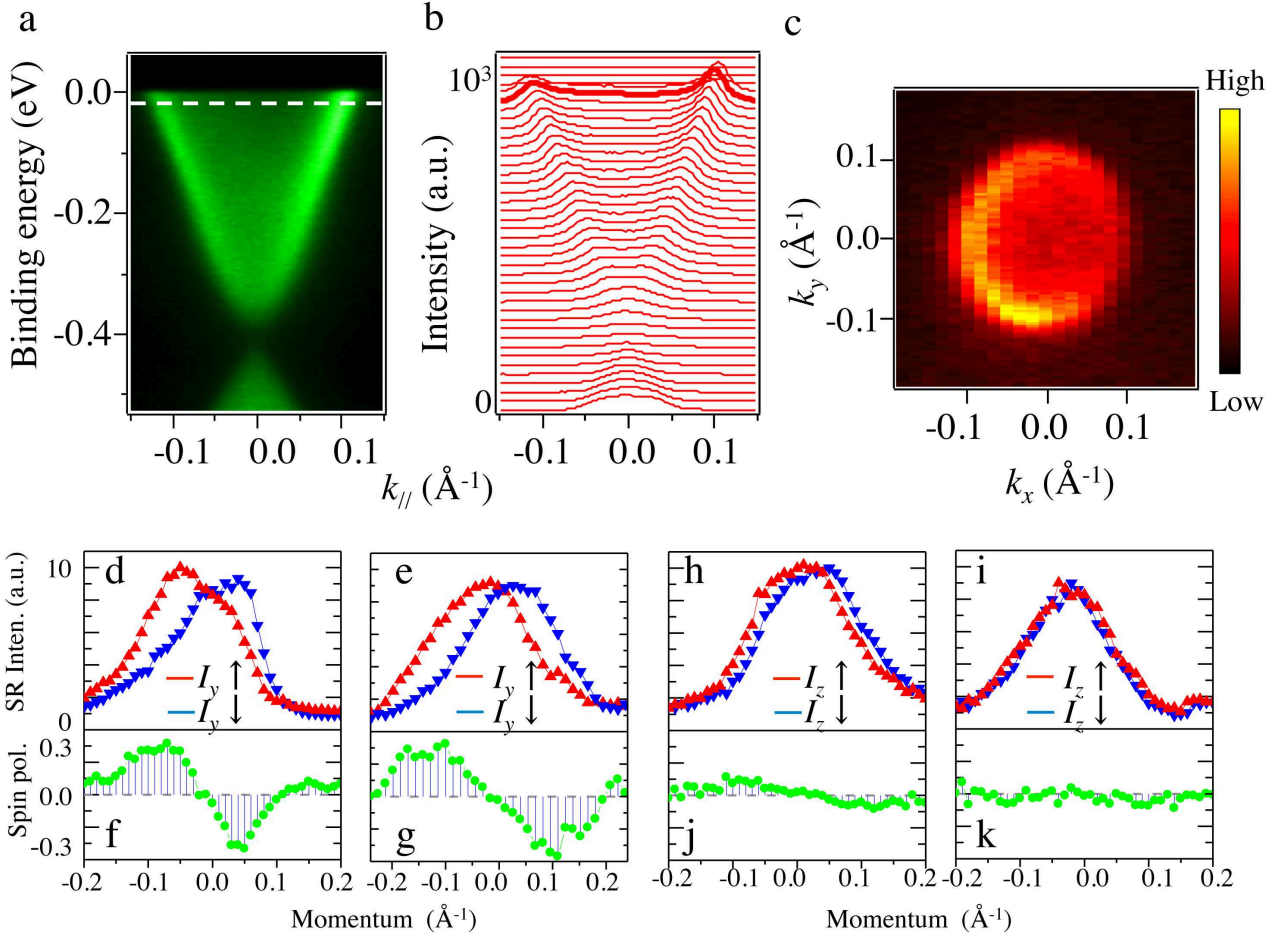}
\caption{\textbf{Observation of spin-momentum locking behavior on the native Fermi level of the gapped surface states in the conventional semiconductor regime.} \textbf{a,} ARPES $k_\|$-$E_{\textrm{B}}$ map of BiTl(S$_{1-\delta}$Se$_\delta$)$_2$ for a $\delta=0.40$ sample. $\delta=0.40$ corresponds to a composition in vicinity of the topological phase transition on the trivial side. Dotted line shows the binding energy where the spin-resolved measurements \textbf{(d-k,)} are performed. \textbf{b,} Momentum distribution curves (MDCs) of the dispersion map in panel \textbf{a}. Highlighted MDC is chosen for spin-resolved measurements. \textbf{c,} Fermi surface mapping for $\delta=0.40$. The two spin-resolved measurements are along the $\bar{\Gamma}$-$\bar{M}$ and $\bar{\Gamma}$-$\bar{K}$ cuts, respectively. Yellow arrows represent the measured spin polarization vectors around the Fermi surface. \textbf{d,e,} measured in-plane spin-resolved (SR) momentum distribution spectra along the $\bar{\Gamma}$-$\bar{M}$ (panel \textbf{d}) and $\bar{\Gamma}$-$\bar{K}$ (panel \textbf{e}) cuts. \textbf{f,g,} measured in-plane net spin polarization (spin pol.) along the along the $\bar{\Gamma}$-$\bar{M}$ (panel \textbf{f}) and $\bar{\Gamma}$-$\bar{K}$ (panel \textbf{g}) cuts. \textbf{h-k,} same as panels \textbf{d-g} but for the out-of-plane component of the spin polarization. $I_y\uparrow$ denote the photoemission intensity whose spin polarization is along the positive direction ($\uparrow$) of the in-plane tangential ($y$) axes.}
\end{figure*}

\begin{figure*}
\centering
\includegraphics[width=17cm]{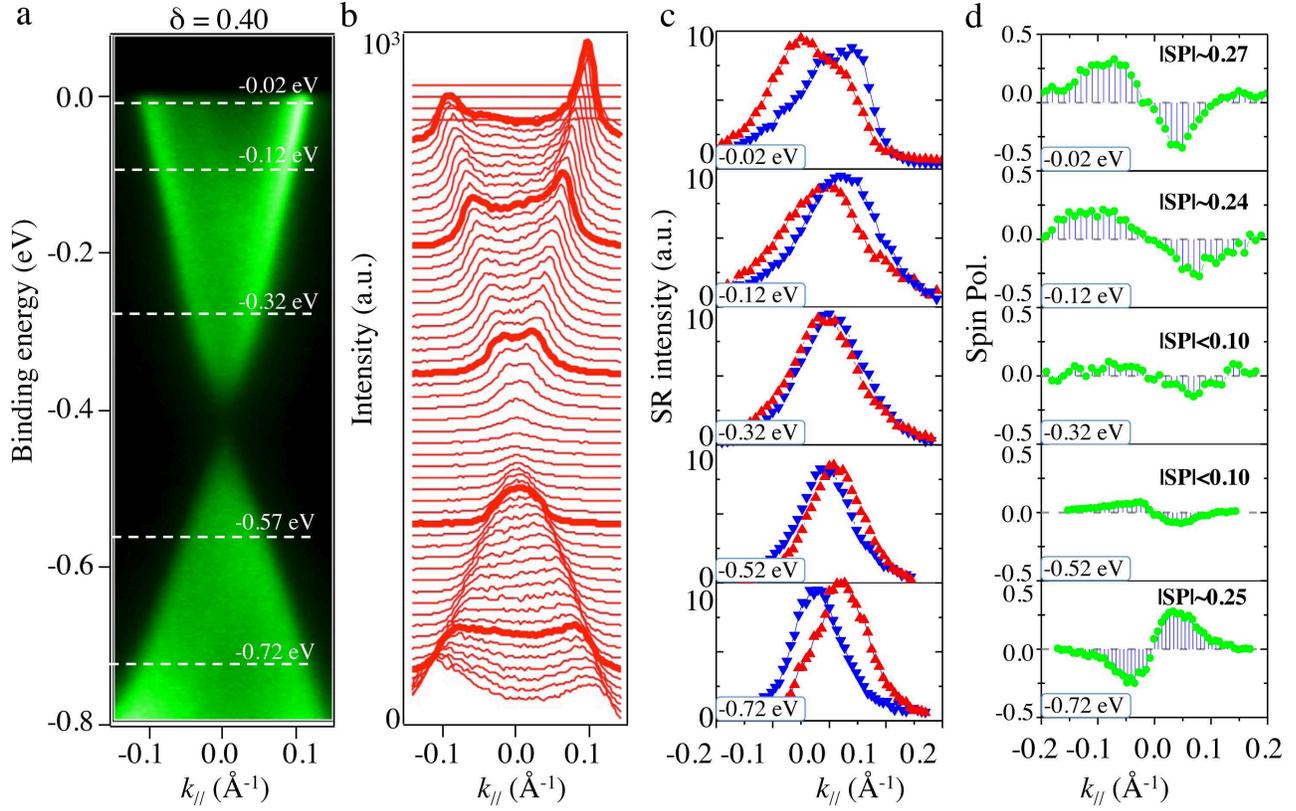}
\caption{\textbf{Evolution of the surface states' spin polarization with binding energy.} \textbf{a,} ARPES $k_\|$-$E_{\textrm{B}}$ map of the $\delta=0.40$ sample with dotted lines indicating the energy levels of spin-resolved measurements. \textbf{b,} MDCs with highlighted curves chosen for spin-resolved measurements. \textbf{c,} Spin-resolved momentum distribution spectra, and \textbf{d,} the corresponding net spin polarization measurements.}
\end{figure*}

\begin{figure*}
\centering
\includegraphics[width=17cm]{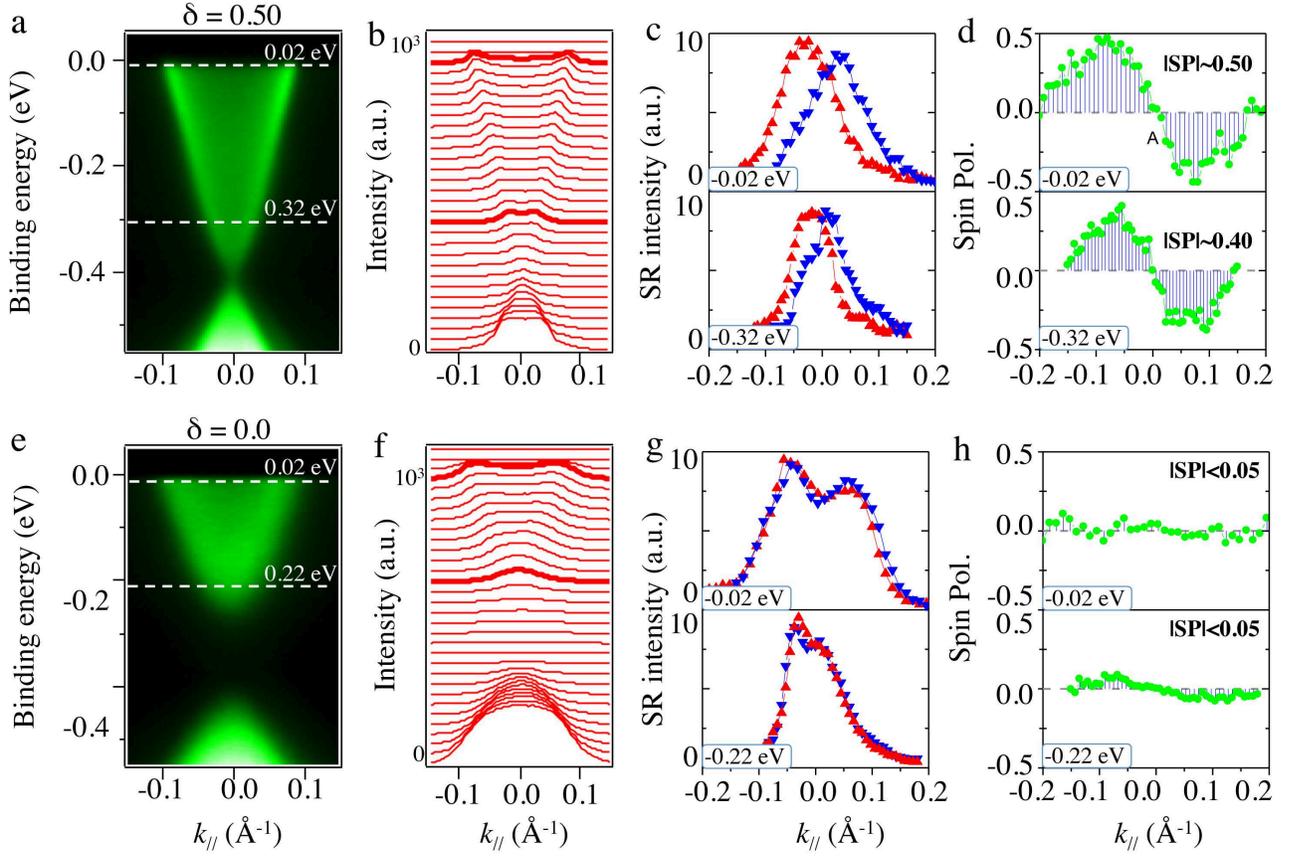}
\caption{\textbf{Evolution of the surface states' spin polarization with composition.} \textbf{a,e,} ARPES $k_\|$-$E_{\textrm{B}}$ maps with dotted lines indicating the energy levels of spin-resolved measurements. Compositions of the samples are marked on the top of each map. \textbf{b,f,} MDCs with highlighted curves chosen for spin-resolved measurements. \textbf{c,g,} Spin-resolved momentum distribution spectra, and \textbf{d,h,} the corresponding net spin polarization measurements.}
\end{figure*}

\clearpage
\begin{figure*}
\centering
\includegraphics[width=17cm]{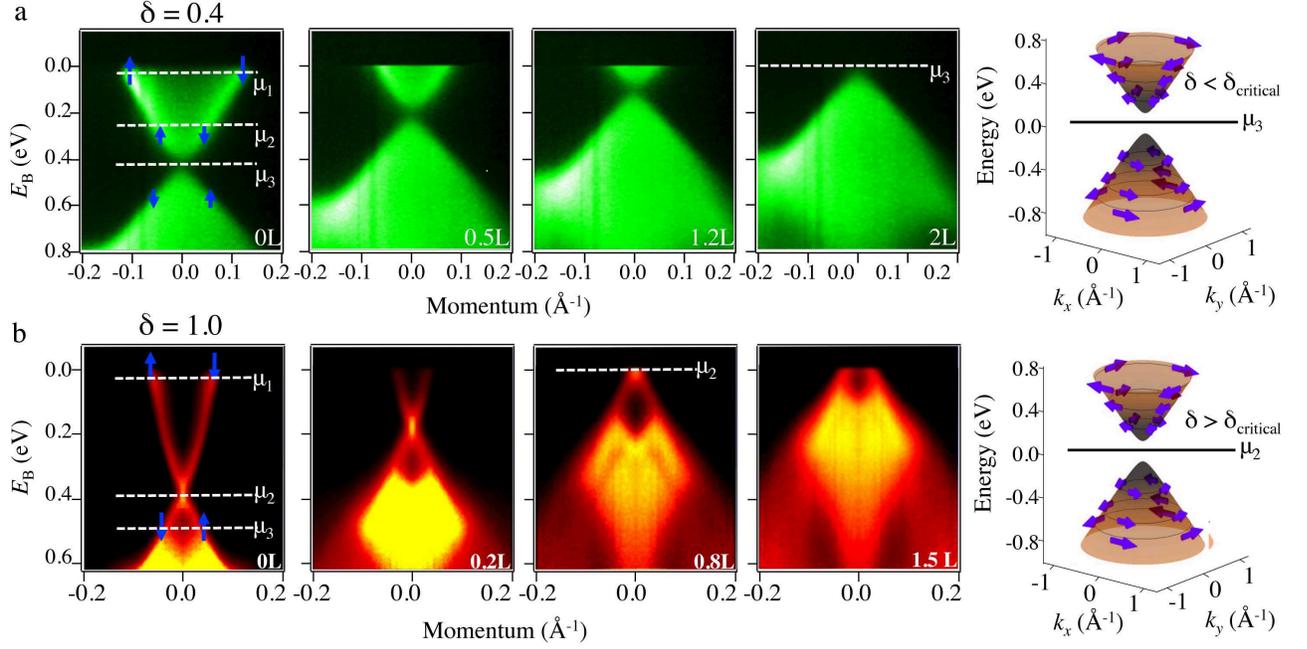}
\caption{\textbf{Spin-momentum locked gapped surface states prior to the topological critical point realize a helical spin switch in a non-topological and non-magnetic setting.} ARPES dispersion measured with incident photon energy of 55 eV under NO$_2$ surface adsorption on the $\delta=0.4$ (panel \textbf{a}) and $\delta=1.0$ (panel \textbf{b}) samples. The NO$_2$ dosage is noted on top of each panel. 1 L $= 1\times10^{-6}$ torr${\cdot}$sec. Blue arrows represent the measured spin polarization of the sample. The length of the arrow qualitatively show the magnitude of the spin polarization. At the chemical potential of $\mu_3$ for the $\delta=0.4$ sample, a surface insulator is realized. Thus the helical spin texture can be switched on and off by tuning the chemical potential.}
\end{figure*}

\end{document}